\documentclass{article}

\usepackage{PRIMEarxiv}
\usepackage[utf8]{inputenc}
\usepackage[T1]{fontenc}
\usepackage{amsmath,amssymb,amsfonts}
\usepackage{graphicx}
\usepackage{placeins}
\usepackage{booktabs}
\usepackage{cite}
\usepackage{hyperref}
\usepackage{url}
\usepackage{microtype}
\usepackage{authblk}
\usepackage{xcolor}
\usepackage{textcomp}
\graphicspath{{figures/}}
\title{Controllable Dysarthric Speech Synthesis with Patient-Specific Conditioning for Speaker-Diverse ASR Augmentation}

\author[1]{Haoshen Wang}
\author[2]{Xueli Zhong}
\author[2]{Bingbing Lin}
\author[2]{Jia Huang}
\author[2,3]{Xingduo Pan}
\author[2]{Shengxiang Liang}
\author[1]{Nizhuan Wang*}
\author[1]{Wai Ting Siok*}

\affil[1]{Department of Language Science and Technology, The Hong Kong Polytechnic University, Hung Hom, Kowloon, Hong Kong SAR, China}
\affil[2]{College of Rehabilitation Medicine, Fujian University of Traditional Chinese Medicine, Fuzhou 350122, China}
\affil[3]{Department of Imaging, Rehabilitation Hospital affiliated to Fujian University of Traditional Chinese Medicine, Fuzhou, Fujian 350003, China}

\begin{document}
\maketitle
\begingroup
\renewcommand{\thefootnote}{*}
\footnotetext{Corresponding authors.}
\endgroup
\begin{abstract}
Dysarthric speech recognition is limited by high speaker variability and scarce labeled data. Existing synthesis methods often couple speaker identity with dysarthric articulation, reducing control over generated speech. We propose a controllable dysarthric speech synthesis framework for ASR augmentation with separate prompt-derived timbre prefixes and learnable patient-specific pathology prefixes. Built on a pre-trained neural codec language model adapted using LoRA, the framework combines both prefixes through additive conditioning. A dual-classifier objective with gradient reversal and voice-conversion-based counterfactual augmentation promotes factor separation, allowing learned pathology conditions to be paired with different target speakers, including healthy speakers. Experiments on TORGO show that generated speech can partially replace real dysarthric training data and provides effective speaker-diverse augmentation when combined with real data. Objective ASR, perceptual, factor-separation, and phoneme-level analyses indicate preservation of target-speaker timbre and pathology-dependent patterns consistent with real dysarthric speech. Code and audio samples are available at \url{https://github.com/Mors20/Controllable-Dysarthric-Speech-Synthesis}.
\end{abstract}


\section{Introduction}

Dysarthria is a motor speech disorder caused by impaired control or weakness of the muscles involved in speech production~\cite{ziegler2003speech}. It commonly produces slurred, slow, or imprecise articulation, hindering communication and challenging automatic speech recognition (ASR) and speech-based assistive technologies. High inter-speaker variability and limited labeled data further complicate dysarthric ASR~\cite{qian2023survey,bhat2025speech,hermann2021handling}. Synthesizing additional dysarthric speech may therefore help alleviate data scarcity and improve downstream ASR robustness~\cite{hermann2023few}.

Existing augmentation includes signal-level perturbations, such as vocal tract length perturbation, tempo stretching, and speed perturbation~\cite{kanda2013elastic,vachhani2018data,geng2022investigation}. Although these methods model speaking-rate and spectral variation, they cannot reproduce perceptual dysarthric characteristics, including articulatory imprecision and atypical voice quality. Generative adversarial networks synthesize disordered speech~\cite{jin2021adversarial,jin2023personalized}, while voice conversion (VC) maps healthy speech to dysarthric realizations. The Rhythm-and-Voice (RnV) framework~\cite{hajal2025unsupervised} uses self-supervised representations to disentangle and convert rhythmic and speaker-related characteristics while adapting ASR with converted speech. DuTa-VC~\cite{wang2023dutavc} introduces duration-aware diffusion VC for typical-to-atypical conversion. Recent prompt-based, large-scale TTS models enable controllable generation~\cite{wang2023neural,guo2023prompttts,shen2024naturalspeech,ju2024naturalspeech,le2023voicebox}. In dysarthric speech synthesis, Soleymanpour et al.~\cite{soleymanpour2024accurate} incorporate a dysarthria severity coefficient and pause-insertion module into neural multi-talker TTS, whereas Leung et al.~\cite{leung2024training} introduce diffusion-based text-to-dysarthric-speech (TTDS) and fine-tune large-scale ASR models with generated data.

\begin{figure*}[t]
  \centering
  \includegraphics[width=0.95\linewidth]{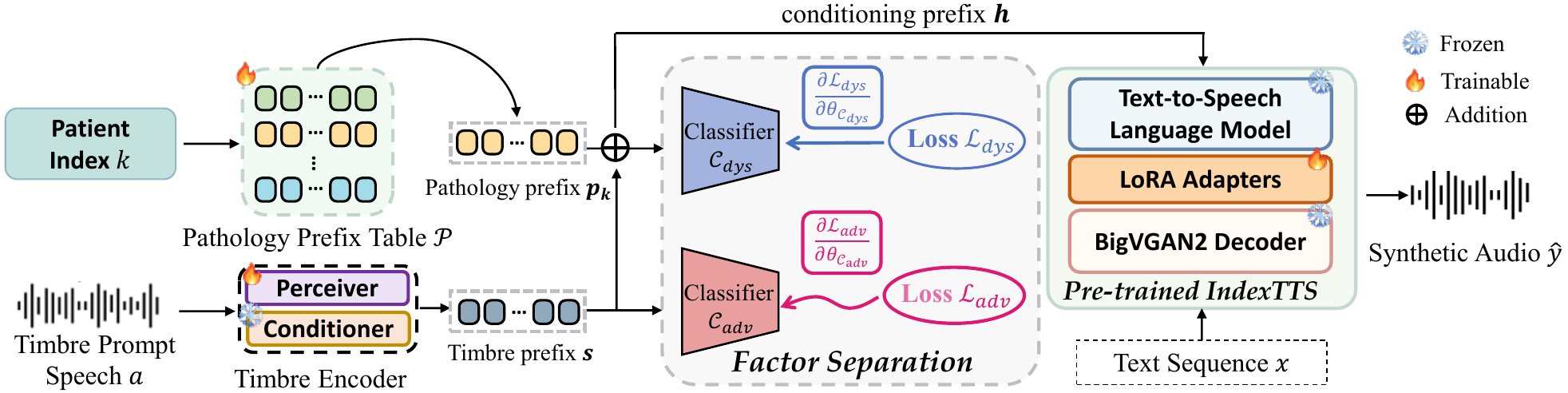}
  \caption{Overview of the proposed controllable dysarthric speech synthesis framework.}
  \label{fig:fig1}
\end{figure*}

Despite these advances, speaker identity and dysarthric articulation can remain coupled during synthesis. Consequently, a patient-specific dysarthric pattern may remain associated with the voice from which it was learned, limiting controllability of augmented data. Promoting separation between speaker timbre and dysarthric articulation can support combining a learned pathology condition with different target voices, as pursued more generally in factorized speech representation learning and multi-factor voice conversion~\cite{qian2020unsupervised,wang2021adversarially}.

In this study, we propose a controllable dysarthric speech synthesis framework with patient-specific conditioning built on a pre-trained IndexTTS backbone~\cite{deng2025indextts}. \textbf{Throughout this paper, pathology denotes a patient-specific dysarthric articulation pattern associated with an observed speaker in the training data, rather than a disease-level clinical subtype, disease etiology, or severity label.} The framework learns one pathology prefix for each observed dysarthric speaker and a shared prefix for the healthy control condition. A timbre encoder extracts a timbre conditioning prefix from a short prompt utterance, which is added to the pathology prefix selected for synthesis.

A dual-classifier objective with a gradient reversal layer (GRL)~\cite{ganin2016domain}, together with VC-based counterfactual augmentation, promotes factor separation between speaker identity and pathology-related attributes. This design supports rendering a learned patient-specific pathology condition with different target prompt speakers, including healthy speakers, for controllable and speaker-diverse ASR augmentation.

The main contributions are summarized as:

\begin{itemize}
    \item We present a controllable dysarthric TTS framework with separately parameterized timbre and pathology conditioning prefixes. Adversarial and counterfactual training promote factor separation and support their cross-combination across observed speakers and pathology conditions.

    \item We show that the resulting synthetic speech provides effective augmentation for dysarthric ASR, serving both as a partial substitute for and an addition to real dysarthric training data.

    \item We assess synthesis behavior through perceptual, representation-level, and phoneme-level analyses, providing evidence on timbre preservation, pathology-dependent control, and agreement between real and synthetic phoneme-error patterns.
\end{itemize}

\section{Method}
\label{sec:method}

\subsection{Problem Formulation}

We formulate controllable dysarthric speech generation using timbre and pathology conditioning. Timbre captures speaker-dependent vocal characteristics, whereas pathology represents patient-specific dysarthric articulation, including atypical prosody, duration, and speech-production patterns.

Given an input text sequence \(x\), a target-speaker prompt audio sample \(a\), and a condition index \(k \in \{0,\dots,n\}\), the goal is to synthesize speech \(\hat{y}\) that preserves the content of \(x\) and the timbre specified by \(a\), while expressing the selected articulation condition. Here, \(k=0\) denotes a shared healthy control condition assigned to all healthy speakers, whereas each \(k \geq 1\) denotes an observed patient-specific dysarthric pathology pattern. The value \(n\) is the total number of patient-specific pathology patterns. An overview is shown in Fig.~\ref{fig:fig1}.

\subsection{Pre-trained TTS Model and Adaptation}

Our proposed method is built on IndexTTS~\cite{deng2025indextts}, a GPT-style neural codec language model that autoregressively predicts a sequence of discrete acoustic codec tokens
\(
\mathbf{c} = (c_1,\dots,c_T)
\)
from text tokens \( x \) and a continuous conditioning prefix
\(
\mathbf{h} \in \mathbb{R}^{L \times D},
\)
where \( L \) denotes the prefix length and \( D \) denotes the dimensionality of the latent embedding space. A neural codec decoder subsequently reconstructs the waveform from the predicted codec tokens.

The pre-trained IndexTTS model is preserved as a strong acoustic prior learned from large-scale healthy speech data. During adaptation, the original model parameters remain frozen except for the perceiver module in the timbre encoder. We further insert low-rank adapters (LoRA)~\cite{LORA} into the attention projections of the text-to-speech language model. This parameter-efficient adaptation allows the model to capture dysarthric speech characteristics from limited data while retaining the synthesis quality and linguistic knowledge of the pre-trained model.

\subsection{Timbre and Pathology Conditioning}

Given a prompt utterance \( a \), the timbre encoder extracts a  timbre conditioning prefix
\(
\mathbf{s} \in \mathbb{R}^{L \times D},
\)
where \( L \) denotes the prefix length and \( D \) denotes the dimensionality of the latent embedding space.

In parallel, we maintain a learnable pathology prefix table
\begin{equation}
\mathcal{P}
=
\{\mathbf{p}_0, \mathbf{p}_1, \dots, \mathbf{p}_n\},
\qquad
\mathbf{p}_i \in \mathbb{R}^{L \times D},
\end{equation}
where \(\mathbf{p}_k\) is selected by condition index \(k\). The prefix
\(
\mathbf{p}_0
\)
represents the shared healthy control condition, while the remaining prefixes correspond to observed patient-specific dysarthric pathology patterns.

For a selected condition index \( k \), the corresponding pathology conditioning prefix \( \mathbf{p}_k \) is combined with the prompt-derived timbre prefix through element-wise addition:
\begin{equation}
\mathbf{h} = \mathbf{s} + \mathbf{p}_k, \qquad \mathbf{s},\,\mathbf{p}_k,\,\mathbf{h} \in \mathbb{R}^{L \times D}.
\end{equation}

The combined conditioning prefix \(\mathbf{h}\) is provided to the IndexTTS language model together with the input text \( x \). This construction allows a selected pathology condition to be combined with timbre information extracted from different target speaker prompts during synthesis.

\subsection{Adversarial Training for Factor Separation}
\label{ssec:factor_sep}

Additive conditioning alone does not ensure that \(\mathbf{s}\) and \(\mathbf{p}_k\) capture different types of information, because pathology-related cues may still be present in the timbre conditioning prefix. We therefore introduce two complementary classification objectives to promote factor separation between timbre-related and pathology-related information.

An adversarial classifier \(\mathcal{C}_{\text{adv}}\) is applied to the timbre conditioning prefix \(\mathbf{s}\) through a gradient-reversal layer (GRL) and trained to distinguish the healthy control condition from dysarthric patient conditions. The GRL is the identity in the forward pass,
\(
\mathrm{GRL}(\mathbf{s})=\mathbf{s},
\)
and negates the gradient on the backward pass,
\(
\partial \mathrm{GRL}(\mathbf{s}) / \partial \mathbf{s}
=
-\lambda \mathbf{I},
\)
where \( \lambda > 0 \) controls the adversarial strength. This objective is intended to reduce pathology-related information in \(\mathbf{s}\).

A condition classifier \(\mathcal{C}_{\text{dys}}\) is then applied to the combined conditioning prefix \(\mathbf{s}+\mathbf{p}_k\) and trained, without gradient reversal, to distinguish the healthy control condition from dysarthric patient conditions. Because the adversarial objective discourages pathology-related information in \(\mathbf{s}\), this condition-prediction objective places additional pressure on \(\mathbf{p}_k\) to retain condition-discriminative information.

Together, two objectives promote the association of pathology-related information with \(\mathbf{p}_k\) rather than \(\mathbf{s}\). The full objective is
\begin{equation}
   \mathcal{L}_{\text{Total}} = \mathcal{L}_{\text{LM}}
   + \alpha\,\mathcal{L}_{\mathcal{C}_{\text{dys}}}
   + \beta\,\mathcal{L}_{\mathcal{C}_{\text{adv}}},
   \label{equation:eq3}
\end{equation}
where \(\mathcal{L}_{\text{LM}}\) is the codec language-model loss and \(\alpha,\beta\) weight the two classifier terms. We linearly ramp \(\lambda\) from \(0\) to its target value over the early training steps to stabilize optimization.

\subsection{Counterfactual Augmentation via Voice Conversion}
\label{ssec:vc-aug}

The adversarial objective in Eq.~\ref{equation:eq3} promotes factor separation between timbre-related and pathology-related information. However, such separation also benefits from training data with sufficient cross-condition variation. In dysarthric corpora, each impaired speaker typically contributes speech in only one voice, creating a strong correlation between dysarthric articulation and speaker timbre.

Thus, we introduce counterfactual augmentation based on zero-shot voice conversion (VC)~\cite{liu2024zero}. Counterfactual augmentation creates training examples in which one factor is changed while the remaining factors are intended to be retained. In our setting, VC changes the speaker timbre of an utterance while aiming to preserve its linguistic content and dysarthric articulation characteristics. Specifically, each dysarthric utterance is converted to the timbre of randomly selected healthy speakers, whereas each healthy utterance is converted to randomly selected dysarthric timbres.

These samples introduce cross-condition variation that is scarce in the original recordings. The same dysarthric articulation source is paired with multiple timbres, while similar timbres occur in both healthy and dysarthric conditions. This variation complements adversarial training and promotes the association of pathology-related information with \(\mathbf{p}_k\) and timbre-related information with \(\mathbf{s}\).

\section{Experiments}

This section first describes the TORGO dataset and leave-one-speaker-out evaluation protocol, followed by four complementary analyses. Objective ASR evaluations assess the utility of generated speech for synthetic-data substitution, real-data augmentation, and comparison with prior TORGO ASR results. Component-wise evaluation quantifies the contributions of adversarial training for factor separation and VC-based counterfactual augmentation using downstream ASR performance and perceptual ranking tasks. Factor-separation analysis characterizes timbre-related and pathology-related variation under controlled synthesis. Finally, phoneme-level error analysis assesses the agreement between recognition patterns in synthetic and real dysarthric speech.

\subsection{Dataset and Evaluation Protocol}
\label{ssec}

The TORGO database of dysarthric speech is used in this study~\cite{rudzicz2012torgo}. It includes recordings from eight English-speaking speakers with dysarthria, comprising three female and five male speakers diagnosed with either cerebral palsy (CP) or amyotrophic lateral sclerosis (ALS), as well as seven healthy control speakers, comprising three female and four male speakers.

The dysarthric speakers were assessed by a speech-language therapist using the Frenchay Dysarthria Assessment (FDA)~\cite{enderby1980frenchay}. Based on these clinical evaluations, the dysarthric speakers are grouped into four severity levels for reporting: Severe (F01, M01, M02, and M04), Moderate-Severe (M05), Moderate (F03), and Mild (F04 and M03). These severity labels are used only for reporting ASR results.

We adopt Whisper models~\cite{radford2023robust} as the ASR backbone and use a leave-one-speaker-out (LOSO) evaluation protocol~\cite{yue2020exploring,leung2024training}.
For each fold, one dysarthric speaker is held out, and that speaker's audio, transcripts, and pathology conditioning prefix are excluded from synthesis and ASR training. Only the remaining speakers are used for training and data generation. The held-out speaker's recordings are used solely for ASR testing, evaluating performance on an unseen target speaker.

\begin{figure}[t]
  \centering
  \includegraphics[width=0.95\linewidth]{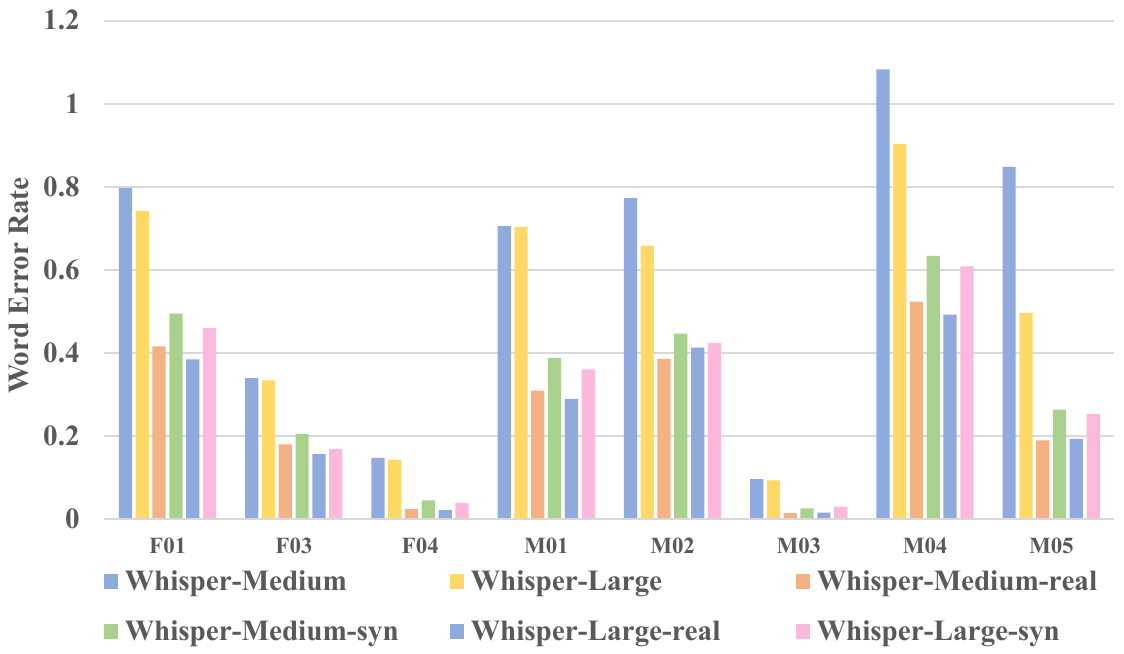}
  \caption{Comparison of ASR performance trained on real and synthetic dysarthric speech.}
  \label{fig:fig2}
\end{figure}

\subsection{Synthetic Data Substitution and Augmentation}
\label{ssec:asr_utility}
We evaluate the utility of the proposed synthetic dysarthric speech for ASR training under two complementary settings: synthetic-data substitution and real-data augmentation. In both settings, we adopt Whisper-Medium and Whisper-Large as ASR backbones and report word error rate (WER) under the LOSO protocol described above.

First, we examine whether synthetic dysarthric speech generated by the proposed framework can serve as a substitute for real dysarthric training data. We compare: (i) the pre-trained Whisper models used without task-specific adaptation, denoted as Whisper-Medium and Whisper-Large; (ii) models fine-tuned using only real dysarthric speech, denoted as Whisper-Medium-real and Whisper-Large-real; and (iii) models fine-tuned using fully synthetic dysarthric speech generated by our framework in place of the real dysarthric training set, denoted as Whisper-Medium-syn and Whisper-Large-syn. This evaluation is conducted across all dysarthric speakers.

The results in Fig.~\ref{fig:fig2} show that fine-tuning with synthetic speech substantially improves ASR performance over the pre-trained models without adaptation. Although models trained on real dysarthric speech achieve the lowest WER overall, the performance gap between real-data and synthetic-data training remains limited. These results indicate that the proposed framework can generate dysarthric speech that provides a useful training signal for dysarthric ASR.

Second, we investigate the benefit of augmenting real training data with increasing amounts of synthetic dysarthric speech. Starting from the real training set, we generate additional dysarthric utterances by combining healthy-speaker prompts with different learned pathology conditioning prefixes. We then augment the real training data with synthetic samples at ratios of 20\%, 40\%, 60\%, 80\%, and 100\%, relative to the size of the real training set. This experiment focuses on the severe dysarthria subset, including F01, M01, M02, and M04, and uses Whisper-Medium as the ASR backbone.

Table~\ref{tab:aug_ratio} reports the resulting WERs. Adding 20\% synthetic speech improves recognition for all evaluated speakers relative to the no-augmentation condition. Although WER is not strictly monotonic across ratios, the lowest WER for each speaker is achieved at either 80\% or 100\% augmentation. Taken together, these results demonstrate that the generated speech can both partially substitute for real dysarthric training data and provide an effective source of speaker-diverse augmentation for dysarthric ASR.

\begin{table}[t]
\centering
\footnotesize
\setlength{\tabcolsep}{4pt}
\caption{Word error rate (WER) under different synthetic data augmentation ratios using Whisper-Medium (WM) on the severe dysarthria subset (F01, M01, M02, M04).}
\label{tab:aug_ratio}
\begin{tabular}{lcccc}
\toprule
Training Setting & F01 & M01 & M02 & M04 \\
\midrule
WM + 0\%   & 0.4161 & 0.3101 & 0.3866 & 0.5244 \\
WM + 20\%  & 0.3080 & 0.2152 & 0.2809 & 0.3682 \\
WM + 40\%  & 0.2866 & 0.1755 & 0.2705 & 0.3078 \\
WM + 60\%  & 0.2619 & 0.1873 & 0.2242 & 0.2719 \\
WM + 80\%  & 0.2553 & 0.1541 & 0.2500 & 0.2591 \\
WM + 100\% & 0.2488 & 0.1535 & 0.2208 & 0.2631 \\
\bottomrule
\end{tabular}
\end{table}

\subsection{Comparison with Prior TORGO ASR Results}
\label{ssec:asr_benchmark}

We further compare our strongest configuration, Whisper-Medium trained on all available real training data and augmented with an equal amount of synthetic speech (100\% augmentation), with prior results reported on the TORGO ASR task, including FS2D~\cite{soleymanpour2024accurate}, FMLLR-DNN~\cite{FMLLR-DNN}, SD-CTL~\cite{SD-CTL}, and TTDS~\cite{leung2024training}. Following prior work, we report mean WER across four severity groups: Severe (F01, M01, M02, and M04), Moderate-Severe (M05), Moderate (F03), and Mild (F04 and M03).

As shown in Table~\ref{tab:benchmark_comparison}, our method achieves the lowest WER in the Severe, Moderate-Severe, and Mild groups. TTDS achieves a slightly lower WER in the Moderate group. These results suggest that speaker-diverse synthetic dysarthric speech is particularly beneficial for the more challenging severity groups, where real training data are limited and acoustic variability is pronounced.

\begin{table}[t]
\centering
\footnotesize
\setlength{\tabcolsep}{3pt}
\caption{WER comparison with prior benchmarks on the TORGO ASR task. }
\label{tab:benchmark_comparison}
\begin{tabular}{lccccc}
\toprule
Method & FS2D & FMLLR-DNN & SD-CTL & TTDS & Ours \\
\midrule
Severe & 0.5588 & 0.4329 & 0.6824 & 0.2330 & \textbf{0.2215} \\
Mod.-Severe & 0.4960 & 0.4405 & 0.3315 & 0.1398 & \textbf{0.1174} \\
Moderate & 0.3680 & 0.3593 & 0.2284 & \textbf{0.0327} & 0.0419 \\
Mild & 0.1260 & 0.1165 & 0.1035 & 0.0257 & \textbf{0.0232} \\
\bottomrule
\end{tabular}
\end{table}

\subsection{Component-wise Evaluation}
\label{ssec:component_eval}

We evaluate adversarial training for factor separation and VC-based counterfactual augmentation from ASR and perceptual perspectives. Study~A measures their contribution to downstream ASR performance, while Study~B examines their association with dysarthria fidelity and timbre preservation.

\paragraph{Study A: ASR impact of design components}
We compare the full model with two variants: w/o Sep. removes the factor-separation objective, including the gradient-reversal layer (GRL) and both auxiliary classifiers; w/o VC removes VC-based counterfactual augmentation while retaining all other components.

Each variant generates the same amount of synthetic dysarthric speech, which augments the same real training set. We fine-tune Whisper-Medium on the resulting datasets and evaluate on held-out real dysarthric speech under the LOSO protocol across the Severe, Moderate-Severe, Moderate, and Mild groups.

Table~\ref{tab:wer_ablation} shows that the full model achieves the lowest WER across groups. Removing either component degrades performance, with the largest degradation from removing factor-separation training in the Severe and Moderate-Severe groups. This indicates its importance for augmentation of more challenging dysarthric speech. Gains are smaller for Mild speech, which Whisper already recognizes relatively well. We perform paired bootstrap resampling over held-out LOSO utterances with 1{,}000 resamples. The full model significantly outperforms both w/o Sep. and w/o VC in pooled WER after Holm correction (\(p < 0.05\)).

\begin{table}[t]
\centering
\footnotesize
\setlength{\tabcolsep}{3pt}
\caption{ASR impact of individual design components. Lower WER is better}
\label{tab:wer_ablation}
\begin{tabular}{lcccc}
\toprule
Method & Severe & Mod.-Severe & Moderate & Mild \\
\midrule
w/o Sep. & 0.3417 & 0.3193 & 0.0617 & 0.0249 \\
w/o VC & 0.2692 & 0.1901 & 0.0533 & 0.0357 \\
Full model & \textbf{0.2215} & \textbf{0.1174} & \textbf{0.0419} & \textbf{0.0232} \\
\bottomrule
\end{tabular}
\end{table}

\paragraph{Study B: Perceptual impact of design components}

The evaluation comprised two ranking tasks. Three senior researchers with expertise in language disorders served as human evaluators. Each evaluator completed 15 trials per task, yielding 45 rank assignments for each condition in each task.

Task~1 evaluates dysarthria fidelity. Human evaluators rank five utterances from most to least dysarthric: real dysarthric speech, a real healthy control, and speech generated by the full model, w/o Sep., and w/o VC.

Task~2 evaluates timbre preservation. Given a dysarthric reference utterance from a patient, human evaluators rank four candidate utterances by perceived similarity to the target prompt speaker's timbre. The candidates include an other-speaker healthy control and synthetic speech generated by the full model, w/o Sep., and w/o VC.

Table~\ref{tab:human_eval} reports the percentage of first-place rankings, top-two rankings, and mean rank. Lower mean rank indicates a stronger match to the target attribute.

In Task~1, real dysarthric speech is ranked most dysarthric in most trials, while the healthy control is consistently ranked last. Among synthetic conditions, the full model obtains the best dysarthria-fidelity result, with a mean rank of 2.00 and a top-two rate of 91.1\%, outperforming both variants.

In Task~2, the full model achieves the best timbre-preservation result, ranking first in 71.1\% of trials and within the top two in 88.9\%. Together with Study~A, these results provide evidence that adversarial training for factor separation and VC-based counterfactual augmentation contribute to ASR utility and perceptual control of the synthesized speech.

\begin{table}[t]
\centering
\small
\caption{Perceptual component-wise evaluation aggregated across three human evaluators. R1 and T2 denote the percentages of trials ranked first and within the top two, respectively. Lower mean rank is better. Boldface indicates the best synthetic system in each task.}
\label{tab:human_eval}
\begin{tabular*}{\columnwidth}{@{\extracolsep{\fill}}lccc@{}}
\toprule
Condition & R1 (\%) & T2 (\%) & Mean $\downarrow$ \\
\midrule
\multicolumn{4}{@{}l}{\textit{Task 1: Dysarthria fidelity}} \\
Real dysarthric & 88.9 & 100.0 & 1.11 \\
Real healthy & 0.0 & 0.0 & 4.96 \\
w/o Sep. & 0.0 & 2.2 & 3.62 \\
w/o VC & 0.0 & 6.7 & 3.31 \\
Full model & \textbf{11.1} & \textbf{91.1} & \textbf{2.00} \\
\midrule
\multicolumn{4}{@{}l}{\textit{Task 2: Timbre preservation}} \\
Other-spk. ctrl. & 0.0 & 0.0 & 3.98 \\
w/o Sep. & 11.1 & 33.3 & 2.58 \\
w/o VC & 17.8 & 77.8 & 2.04 \\
Full model & \textbf{71.1} & \textbf{88.9} & \textbf{1.40} \\
\bottomrule
\end{tabular*}
\end{table}

\begin{figure}[t]
\centering
\includegraphics[width=0.95\columnwidth]{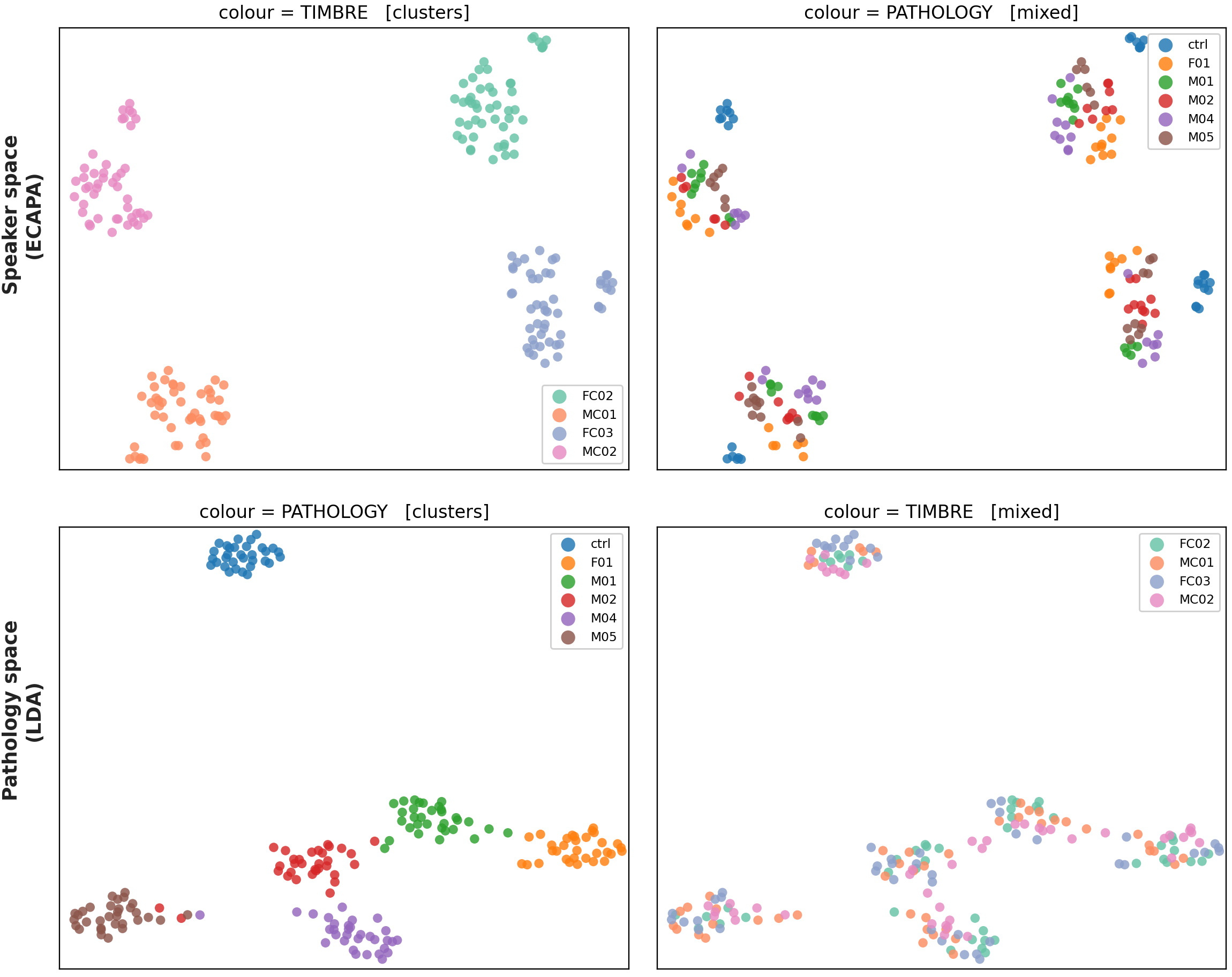}
\caption{Factor separation analysis of timbre and pathology in generated speech, visualized using t-SNE on held-out utterances. \textbf{Top:} The speaker space clusters primarily by timbre condition, while pathology conditions are mixed within clusters. \textbf{Bottom:} The pathology space clusters by pathology condition, while target timbres are mixed within clusters. }
\label{fig}
\end{figure}
\begin{figure}[t]
\centering
\includegraphics[width=0.85\columnwidth]{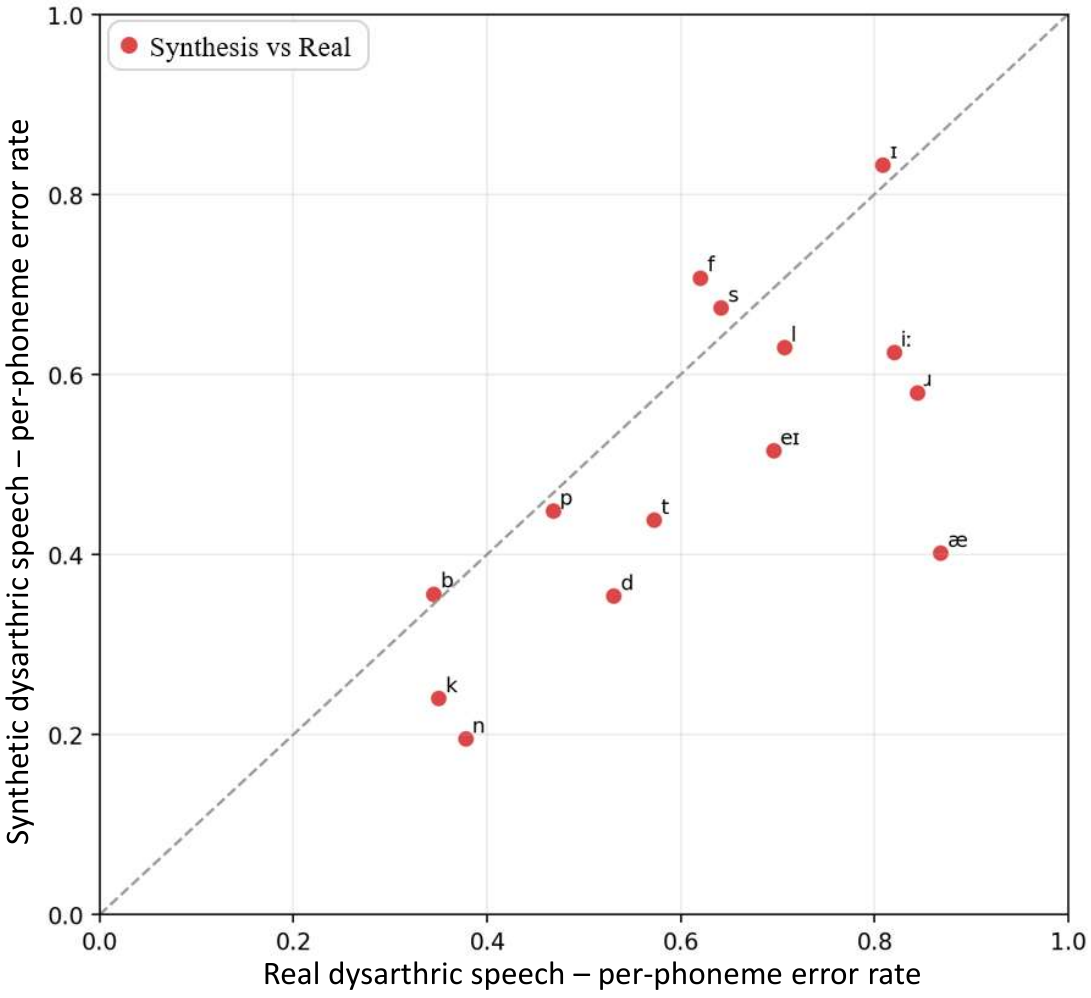}
\caption{Comparison of phoneme-level error rates between real and synthetic dysarthric speech. Each point represents one phoneme. The horizontal axis shows the error rate measured on real dysarthric speech, while the vertical axis shows the corresponding error rate measured on synthetic dysarthric speech. The dashed diagonal indicates exact agreement between the two profiles.}
\label{fig:phoneme_pattern}
\end{figure}
\begin{figure}[t]
\centering
\includegraphics[width=0.85\columnwidth]{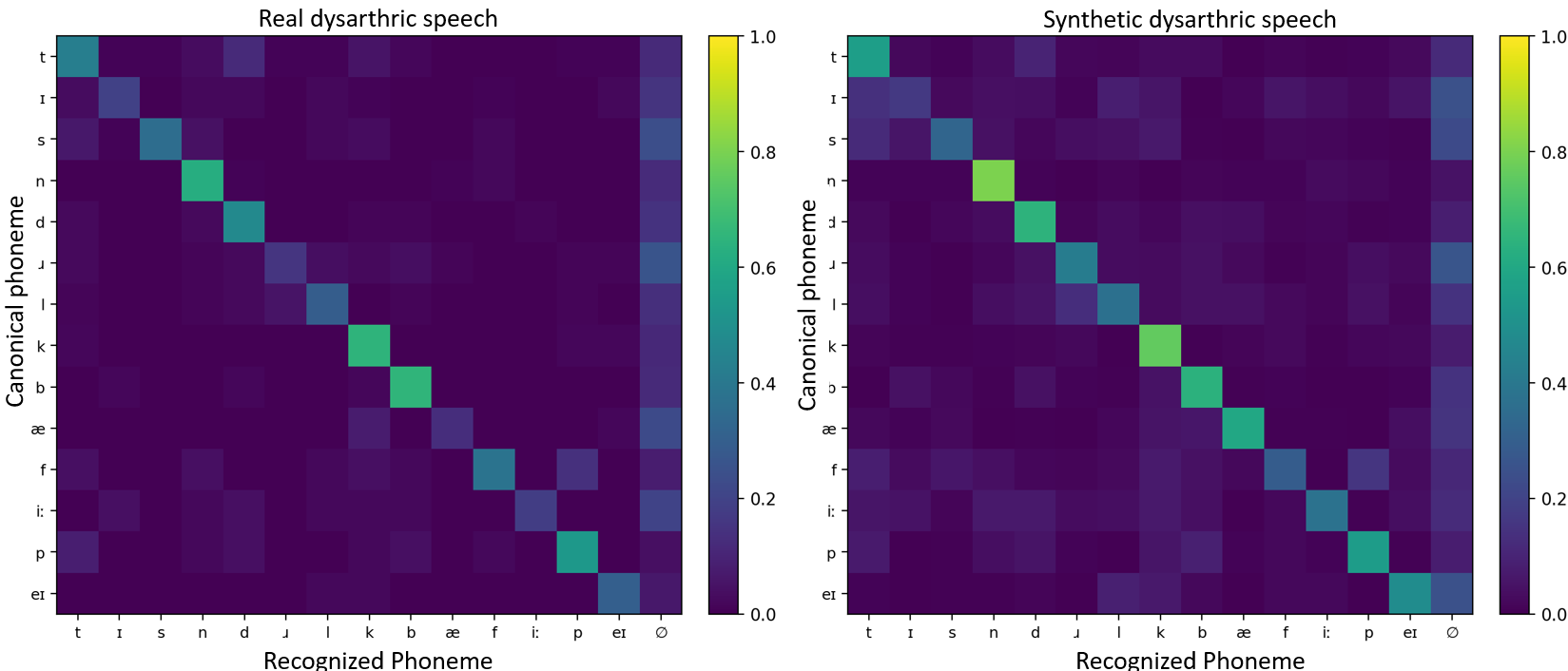}
\caption{Row-normalized phoneme confusion matrices for real and synthetic dysarthric speech. Rows denote canonical phonemes and columns denote recognized phonemes. The symbol \(\varnothing\) denotes a deletion. Both matrices use the same phoneme ordering and color scale.}
\label{fig:phoneme_confusion}
\end{figure}

\subsection{Factor Separation Analysis}
\label{ssec:factor_separation_analysis}

We assess whether the proposed training strategy promotes distinct timbre- and pathology-related variation under controlled synthesis. We synthesize all \( \text{pathology} \times \text{timbre} \) combinations for five TORGO speakers with severe or moderate-severe dysarthria and one healthy control condition. These speakers are selected because their pronounced articulation deficits provide clearer pathology-dependent variation. Each condition is rendered with four healthy target-speaker timbres over fixed sentences. The synthesized utterances are evaluated in speaker and condition-discriminative representation spaces to assess timbre- and pathology-related variation, respectively.

The speaker space consists of ECAPA-TDNN speaker embeddings~\cite{desplanques2020ecapa}, which are designed to capture voice identity.
The pathology space is constructed from a condition-discriminative utterance-level representation extracted by a frozen classifier trained to distinguish the six pathology conditions. Specifically, we use the pooled hidden representation before its classification layer. These representations are projected onto the discriminant axes of an LDA model fitted on a stratified 50\% split, while visualizations are computed on the held-out 50\%.
Both representation spaces are visualized in two dimensions using t-SNE.

Fig.~\ref{fig} summarizes the results. In the speaker space (top row), the embeddings form four compact clusters corresponding to the four target timbre conditions. When the same embeddings are colored by pathology, samples from different pathology conditions are intermingled within each timbre cluster. This pattern is consistent with the timbre prompt exerting the dominant effect on speaker-space variation in the generated speech.

The pathology space (bottom row) shows the complementary pattern. The generated utterances form clusters associated with the six pathology conditions, while samples rendered with different speaker timbres are mixed within each pathology cluster. Within this controlled analysis, these results indicate that pathology conditions are recoverable in the pathology space with limited dependence on the evaluated target timbres.

Taken together, the speaker-space and pathology-space analyses provide evidence consistent with factor separation in the generated speech. They do not establish complete independence between timbre and pathology, but suggest that the proposed training strategy promotes their association with different representation spaces under controlled synthesis.

\subsection{Phoneme-error Pattern Fidelity}
\label{ssec:phoneme_pattern}

Beyond aggregate ASR performance, we assess the agreement between phoneme-level recognition patterns in real and synthetic dysarthric speech. For each canonical phoneme, we align the reference phoneme sequence with the sequence recognized from real or synthetic speech and compute the corresponding phoneme error rate. Synthetic utterances are generated using the pathology conditioning prefixes associated with the corresponding dysarthric speakers. We then compare the resulting phoneme-level error profiles and quantify their agreement using the Pearson correlation coefficient.

Fig.~\ref{fig:phoneme_pattern} compares phoneme error rates measured on synthetic and real dysarthric speech. The two profiles show a positive Pearson correlation of \(r=0.69 \), indicating that phonemes with relatively high error rates in real speech also tend to be more difficult to recognize in the synthesized speech. Several highly error-prone phonemes lie close to the diagonal, whereas others deviate from the equality line. Such deviations are expected because synthesized speech does not reproduce every patient-specific acoustic realization, speaking habit, or recording characteristic present in the real data.

To further characterize recognition errors, we compute phoneme confusion matrices for real and synthetic dysarthric speech. Each matrix is normalized by canonical phoneme, such that each row represents the distribution of recognized phonemes conditioned on a reference phoneme. Diagonal entries reflect correct recognition, while off-diagonal entries capture substitutions. The symbol \(\varnothing\) denotes deletion errors.

Fig.~\ref{fig:phoneme_confusion} shows that the real and synthetic confusion matrices exhibit similar recognition structures. In both cases, some phonemes retain strong diagonal entries, whereas others show lower correct-recognition probabilities and greater confusion with alternative phonemes or deletion. The synthetic matrix also shows elevated deletion tendencies for several phonemes that are similarly error-prone in real speech.

Taken together, the error-rate comparison and confusion-matrix analysis provide descriptive evidence that the synthetic speech reflects aspects of the phoneme-specific recognition patterns observed in real dysarthric speech. The error-rate comparison captures relative phoneme difficulty, while the confusion matrices characterize the distribution of substitutions and deletions.

\FloatBarrier

\section{Conclusion}

In this paper, we presented a controllable dysarthric speech synthesis framework with patient-specific conditioning for ASR augmentation. It combines a prompt-derived timbre prefix with a learnable pathology prefix. Adversarial training and VC-based counterfactual augmentation promote separation of timbre- and pathology-related information, enabling learned pathology conditions to be rendered with different target speaker timbres.

Experiments on TORGO show that generated speech can partially substitute for real dysarthric training data and effectively augment it. Component-wise evaluations show that adversarial training and VC-based augmentation improve ASR performance, dysarthria fidelity, and timbre preservation. Representation and phoneme-level analyses provide evidence consistent with factor separation and agreement with recognition patterns in real dysarthric speech. These findings suggest patient-specific conditioning as a practical approach to expanding dysarthric ASR training data.

\section*{Acknowledgments}
This work was supported by The Hong Kong Polytechnic University Start-up Fund (Project ID: P0053210), The Hong Kong Polytechnic University Faculty Reserve Fund (Project ID: P0053738), an internal grant from The Hong Kong Polytechnic University (Project ID: P0048377), The Hong Kong Polytechnic University Departmental Collaborative Research Fund (Project ID: P0056428), The Hong Kong Polytechnic University Collaborative Research with World-leading Research Groups Fund (Project ID: P0058097) and Research Grants Council Collaborative Research Fund (Ref: C5033-24G).

\bibliographystyle{unsrt}
\bibliography{main}

\end{document}